\documentclass[pra,onecolumn,showpacs]{revtex4}
\usepackage{graphicx}
\usepackage{dcolumn}
\usepackage{bm}
\usepackage{amssymb}
\usepackage{amsmath}
\usepackage{epsfig}

\begin{document}

\title{Experimental Measurement of Multi-dimensional Entanglement via
Equivalent Symmetric Projection}
\author{F. W. Sun}
\email{fwsun@mail.ustc.edu.cn}
\author{J. M. Cai}
\author{J. S. Xu}
\author{G. Chen}
\author{B. H. Liu}
\author{C. F. Li}
\author{Z. W. Zhou}
\email{zwzhou@ustc.edu.cn}
\author{G. C. Guo}
\affiliation{Key Laboratory of Quantum Information, University of Science and Technology
of China, CAS, Hefei, 230026, the People's Republic of China }
\date{\today }

\begin{abstract}
We construct a linear optics measurement process to determine the
entanglement measure, named \emph{I-concurrence}, of a set of $4
\times 4$ dimensional two-photon entangled pure states produced in
the optical parametric down conversion process. In our experiment,
an \emph{equivalent} symmetric projection for the two-fold copy of
single subsystem (presented by L. Aolita and F. Mintert, Phys.
Rev. Lett. \textbf{97}, 050501 (2006)) can be realized by
observing the one-side two-photon coincidence without any
triggering detection on the other subsystem. Here, for the first
time, we realize the measurement for entanglement contained in
bi-photon pure states by taking advantage of the
indistinguishability and the bunching effect of photons. Our
method can determine the \emph{I-concurrence} of generic high
dimensional bipartite pure states produced in parametric down
conversion process.
\end{abstract}

\pacs{03.65.Ud, 03.67.Mn, 42.50.Dv}
\maketitle

\section{Introduction}
The characterization and quantification for quantum entanglement
has become one of the most central issues in quantum information
theory. Various approaches for characterization of entanglement in
quantum states have been proposed. These are based on quantum
state tomography \cite {James,Thew,Roos}, entanglement witnesses
\cite {Bourennane,Haffner,Walther,Lu,Kiesel,Leibfried}, and
quantum nonlocality\cite {Pan,Eibl,Howell,Thew1}. The most
important approach for quantifying the amount of entanglement in
any quantum state is entanglement measure. Up till now, several
well defined entanglement measures have been established, e.g.
concurrence \cite{ConC} and entanglement of formation \cite{EOF}.
The experimental determination of entanglement measure is,
however, a very difficult task, since many measures are
complicated nonlinear functions of the density matrix of the
quantum state. The situation is even worse for multipartite and
multi-dimensional quantum systems. The most straightforward way is
to reconstruct the quantum state fully through quantum state
tomography \cite {James,Thew,Roos}. However, this method has the
disadvantage in being unscalable, and not all the state parameters
are necessary for the determination of entanglement measures.

Recently, there are increasing interests in the entanglement
measured by concurrence, originally defined for two-qubit
entanglement and later generalized to multipartite and
multi-dimensional quantum systems
\cite{I-concurrence,Mintert04PRL,Mintert05PRL,Mintert06PRL}. It is
one of the most fundamental entanglement measures and has been
widely used in many fields of quantum information theory, e.g. the
research of entanglement in quantum phase transitions
\cite{Entanglement and Phase Transitions}. One important property
of concurrence is that it depends on a polynomial function of the
elements of the density matrix. This makes it possible to observe
concurrence through some appropriate observables with two-fold
copy of quantum states \cite{Mintert06PRL}. In Ref.\cite{Walborn},
Walborn, \emph{etc.} reported an experimental determination of
concurrence for two-qubit pure states.

In this paper, we report an experimental determination of the
generalized concurrence \cite{I-concurrence}, i.e.
\emph{I-concurrence}, of $4\times 4$ dimensional pure states
produced by optical parametric down conversion (PDC) by using the
polarization and time-energy mode. Different from the measurements
on two-fold copy of quantum states, here, our strategy is to
detect coincidence counts of high order optical PDC directly,
which contains all the information about the amplitudes in low
order optical PDC and corresponds to an \emph{equivalent}
symmetric projection for the two-fold copy of single subsystem.
Our scheme is simpler than the two-fold copy measurement,
moreover, can be generalized to the measurement of higher
dimensional bipartite pure states produced in optical PDC.

The structure of this paper is as follows. In Sec. II we present a
brief review for \emph{I-concurrence} of bipartite pure states. In
Sec. III quantum states produced in optical PDC are investigated.
We find the \emph{I-concurrence} of an entangled state produced in
1-order optical PDC can be measured by detecting 2-order optical
PDC process, which corresponds to the implementation of an
equivalent symmetric projection for two-fold copy of this state.
Sec. IV depicts an experimentally implementation for the
equivalent symmetric projection for two-fold copies of a set of $4
\times 4$ dimensional two-photon entangled pure states. Sec. V
contains conclusions and some discussions.

\section{A Review for I-Concurrence}
For a pure state $|\psi \rangle $ of a $d_{1}\times d_{2}$ quantum
system, the \emph{I-concurrence} is defined as
\cite{I-concurrence}

\begin{equation}
C=\sqrt{2(1-Tr\rho _{1}^{2})}
\end{equation}%
where $\rho _{1}$ is the reduced density matrix of the $1$st
subsystem. The above generalized concurrence is simply related to
the purity of the marginal density matrices. The maximum value of
\emph{I-concurrence} is $\sqrt{ 2(M-1)/M}$, where $M=\min
(d_{1},d_{2})$. We note that $Tr\rho _{1}^{2}$ is a quadratic
function of the elements of the density matrix $\rho_{1}$. Thus,
one could always find an observable $\hat{A}$ on $2$ copies of
$\rho _{1}$, such that $Tr\rho _{1}^{2}=Tr( \hat{A}\rho
_{1}\otimes \rho _{1})$ \cite{MPFS}. This allows to measure $C$
without quantum state tomography. Actually, it also has been shown
that $Tr\rho _{1}^{2}=1-2Tr(P_{-}^{1}\rho _{1}\otimes \rho
_{1})=2Tr(P_{+}^{1}\rho _{1}\otimes \rho _{1})-1$, where
$P_{+}^{1}$ and $ P_{-}^{1}$ are the projectors onto the symmetric
and antisymmetric subspace of the Hilbert space
$\mathcal{H}_{1}\otimes \mathcal{H}_{1}$, which describes the
two-fold copy of the $1$st subsystem \cite{Mintert06PRL}.
Therefore, the \emph{I-concurrence} can be expressed as the
expectation value of a Hermitian operator $\hat{A}$ on
$\mathcal{H}\otimes \mathcal{H}$, i.e.
\begin{equation}
C=\sqrt{\langle \psi |\otimes \langle \psi |\hat{A}|\psi \rangle \otimes
|\psi \rangle }
\end{equation}%
where $\hat{A}=4P_{-}^{1}=4(I-P_{+}^{1})$. Thus, we can determine
the \emph{I-concurrence} by measuring one single factorizable
observable $\hat{A}$ on two-fold copy of one subsystem.

\section{I-Concurrence for two-photon states produced in Optical PDC}
Experimentally, entangled two-photon state can be produced through
optical PDC. When we consider different degrees of freedom (DOF)
of photons, such as polarization, time-energy, \textit{etc}, high
dimensional entangled pure states can be constructed by using
appropriate linear optical methods. In the Schimdt decomposition,
the high dimensional bipartite pure state can be represented as: $
\left\vert \Psi _{2}\right\rangle =\sum_{i}\sqrt{\lambda
_{i}}\left\vert A_{i}\right\rangle \left\vert B_{i}\right\rangle
$, with $\sum_{i}\lambda _{i}=1$, or $\left\vert \Psi
_{2}\right\rangle =\sum_{i}\sqrt{\lambda _{i}}a_{i}^{\dag
}b_{i}^{\dag }\left\vert vac\right\rangle$, where $a_{i}^{\dag }$
and $b_{i}^{\dag }$ are the $ith$ mode photon creation operators
on the side of $A$ and $B$, respectively. However, indeed, a fully
representation for the state produced in optical PDC is
\cite{Lamas}:
\begin{equation}
\left\vert \Psi \right\rangle =\left\vert vac\right\rangle
+\sqrt{\eta } \left\vert \Psi _{2}\right\rangle +\frac{\eta
}{2!}\left\vert \Psi _{4}\right\rangle +...\text{,}
\end{equation}%
where $\left\vert \Psi _{4}\right\rangle $ refers to the
four-photon state, which has the form: $\left\vert \Psi
_{4}\right\rangle =\sum_{ij}\sqrt{\lambda _{i}\lambda _{j}}
a_{i}^{\dag }a_{j}^{\dag }b_{i}^{\dag }b_{j}^{\dag }\left\vert
vac\right\rangle$. In general, due to the amplitude of the
probability $\left\vert \eta \right\vert \ll 1$, the effect from
the multi-photon components can be omitted only when the behavior
of two-photon is investigated. But, here, we find an easy way to
measure the \emph{I-concurrence} of the state $\left\vert \Psi
_{2}\right\rangle $ by taking advantage of partially probing
four-photon component $\left\vert \Psi _{4}\right\rangle $.

It should be noted that $\left\vert \Psi_{4}\right\rangle $ is not
the product of two two-photon state $\left| \Psi _2\right\rangle
\left| \Psi _2\right\rangle =\sum_{ij}\sqrt{\lambda _i\lambda
_j}a_i^{\dag }a_j^{^{\prime }{\dag }}b_i^{\dag }b_j^{^{\prime
}{\dag }}\left| vac\right\rangle $ (where $a_i^{\dag }$and
$a_j^{^{\prime }{\dag }}(b_i^{\dag }$ and $b_j^{^{\prime }{\dag
}})$ refer to creation operators for different modes) despite they
have very similar forms. We name $\left\vert \Psi
_{4}\right\rangle $ pseudo two-fold copy of state $\left\vert \Psi
_{2}\right\rangle $. But, there is an intriguing relation between
state $\left\vert \Psi _{4}\right\rangle $ and state $\left| \Psi
_2\right\rangle \left| \Psi _2\right\rangle$:
\begin{equation}
\left\langle \Psi _2\right| \left\langle \Psi _2\right|
4P_{+}^{1}\left| \Psi _2\right\rangle \left| \Psi _2\right\rangle
=\left\langle \Psi _4|\Psi _4\right\rangle=2(1+\sum_i\lambda _i^2)
\text{.}
\end{equation}
Here, The projector $P_{+}^1$ can be represented as:
$\sum_{i<j}\frac 12\left[ \left( a_i^{\dag }a_j^{^{\prime}{\dag
}}+a_j^{\dag }a_i^{^{\prime }{\dag }}\right) \left|
vac\right\rangle \left\langle vac\right| \left( a_ia_j^{^{\prime
}}+a_ja_i^{^{\prime }}\right) \right] +\sum_ia_i^{\dag
}a_i^{^{\prime }{\dag }}\left| vac\right\rangle \left\langle
vac\right| a_ia_i^{^{\prime }}$. The four-photon state $\left|
\Psi _4\right\rangle $ can be rewritten as:
\begin{equation}
\left| \Psi _4\right\rangle =\sum_{i<j}2\sqrt{\lambda _i\lambda
_j}a_{i}^{\dag }a_{j}^{\dag }b_{i}^{\dag }b_{j}^{\dag }\left\vert
vac\right\rangle+\sum_i\lambda _ia_i^{\dag2}b_i^{\dag2}\left|
vac\right\rangle \text{,}
\end{equation}%
where the first item indicates that $a_{i}^{\dag }a_{j}^{\dag
}b_{i}^{\dag }b_{j}^{\dag }\left\vert vac\right\rangle$ and
$a_{j}^{\dag }a_{i}^{\dag }b_{j}^{\dag }b_{i}^{\dag }\left\vert
vac\right\rangle$ are not distinguishable, and the second item
will cause the bunching effect of identical photons. So, we may
deduce that: $\left\langle \Psi _2\right| \left\langle \Psi
_2\right| 4P_{+}^{1}\left| \Psi _2\right\rangle \left| \Psi
_2\right\rangle =\left\langle \Psi _4|\Psi
_4\right\rangle=4(\sum_{i<j}\lambda _i\lambda _j+\sum_i\lambda
_i^2)=2(1+\sum_i\lambda _i^2) \text{.}$

\begin{figure}[tbh]
\begin{center}
\includegraphics[width= 2in]{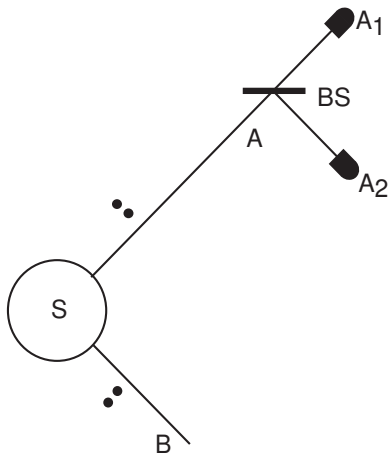}
\end{center}
\caption{ Equivalent symmetric projection measurement for the
two-fold copy of one subsystem is achieved from the coincidence
counts between $A_{1}$ and $A_{2}$ .} \label{fig1}
\end{figure}

Eq.(4) suggests we can determine the entanglement of state
$\left\vert \Psi _{2}\right\rangle $ by probing the inner product
of the state $\left\vert \Psi _{4}\right\rangle $, which can be
realized by counting two-photon coincidence between the two output
ports of the unpolarized symmetric beamsplitter for photons from a
single optical PDC source. Calculation shows that the contribution
for coincidence count between $A_{1}$ and $A_{2}$ (see Fig.1) is
solely from the component $\left\vert \Psi _{4}\right\rangle $ in
the full wavefunction $\left\vert \Psi \right\rangle $ in Eq.(3).
(Here, we omit multi-photon $(\geqslant 6)$ components due to
their tiny probability.) When mode A passes through the
beamsplitter (see Fig.1), $\left\vert \Psi _{4}\right\rangle $ is
transformed into the superposition of two orthogonal components:
$\frac 1{\sqrt{2}}(\left\vert \Psi _{c}\right\rangle + \left\vert
\Psi _{s}\right\rangle)$, where $\left\vert \Psi _{c}\right\rangle
$ is the component of wavefunction giving rise to coincidence
counts between $A_{1}$ and $A_{2}$ and $\left\vert \Psi
_{s}\right\rangle $ is not. The form of $\left\vert \Psi
_{c}\right\rangle $ is as follows:
\begin{eqnarray}
\left\vert \Psi _{c}\right\rangle &=&[\sum_{i<j}\sqrt{\lambda
_{i}\lambda _{j}}(a_{i,A_{1}}^{\dag }a_{j,A_{2}}^{\dag
}+a_{j,A_{1}}^{\dag
}a_{i,A_{2}}^{\dag })b_{i}^{\dag }b_{j}^{\dag }  \notag \\
&&+\sum_{i}\lambda _{i}a_{i,A_{1}}^{\dag }a_{i,A_{2}}^{\dag }b_{i}^{\dag
2}]\left\vert vac\right\rangle \text{,}
\end{eqnarray}%
Then the probability of the two-photon counts after the BS is:
$P_{A_{1}A_{2}}=\frac{1}{4}\eta _{_{A_{1}}}\eta
_{_{A_{2}}}\left\vert \eta \right\vert ^{2}\left\langle \Psi
_c|\Psi _c\right\rangle=\frac{1}{8}\eta _{_{A_{1}}}\eta
_{_{A_{2}}}\left\vert \eta \right\vert ^{2}\left\langle \Psi
_4|\Psi _4\right\rangle=P_{A_{1}}P_{A_{2}}(\sum_{i}\lambda
_{i}^{2}+1)$, where $P_{A_{1}(A_{2})}=\frac{1}{2}\eta
_{_{A_{1}(A_{2})}}\left\vert \eta \right\vert$ is the single
photon counts probability \cite{efficiency}. $\eta _{_{A_{1}}}$
and $\eta _{_{A_{2}}}$ are photon collection efficiencies
including the effect of photon coupling losses and the detector
efficiency. Here, we find that the probability of the two-photon
coincidence counts $P_{A_{1}A_{2}}$ is always larger than the
product of single photon counts probabilities $P_{A_{1}}$ and
$P_{A_{2}}$. The reason relies on the indistinguishability and the
bunching effect of photons in $\left| \Psi _4\right\rangle$. By
defining $K=Tr\rho _{1}^{2}=\frac{
P_{A_{1}A_{2}}}{P_{A_{1}}P_{A_{2}}}-1$, the \emph{I-concurrence}
of a bipartite pure state is
\begin{equation}
C=\sqrt{2-2K}\text{.}
\end{equation}

\section{Experiment}
In our experiment, polarization and time-energy DOF are used to
realize a 4 dimensional Hilbert space. Different polarization
states are produced with two type-I PDC \cite{Kwait} and
time-energy DOF are introduced by using a 52.4 mm quartz crystal
(QC) to induce different time delays for two polarization
components of the pump beam. This is shown in Fig.2. The half wave
plate (HWP) before the QC rotates the pump beam polarization.
After the QC, the pump beam state is:
\begin{equation}
\left\vert P\right\rangle =\cos 2\theta _{1}\left\vert HT_{1}\right\rangle
+\sin 2\theta _{1}\left\vert VT_{2}\right\rangle \text{,}
\end{equation}%
where $\theta _{1}$ is the angle of HWP$_{1}$ and the time delay
$\Delta T=|T_{1}-T_{2}|=$1.68 ps. The pulse laser beam with a
pulse width of $\tau _{p}$=150 fs and repetition rate of $f$=76
MHz from a Ti: Sapphire ultra-fast laser (Coherent D-900) is
frequency doubled to 390 nm, which serves as the pump beam to two
degenerated noncollinear type-I cut BBOs with mutually
orthogonally optical axes. Each $\left\vert T_{i}\right\rangle$
($i=1$, $2$) pulse generates a two-photon entangled state if we
adjust the angle $\theta _{2}$ of HWP$_{2}$ \cite {Kwait}.
Superposition of the two entangled polarization states with
different $T_{i}$ is a two-photon four-dimensional state:
\begin{eqnarray}
\left\vert \Psi _{2}\right\rangle  &=&\cos 2\theta _{1}(\cos 2\theta
_{2}\left\vert VT_{1}\right\rangle \left\vert VT_{1}\right\rangle +\sin
2\theta _{2}\left\vert HT_{1}\right\rangle \left\vert HT_{1}\right\rangle )
\notag \\
&&+\sin 2\theta _{1}(\sin 2\theta _{2}\left\vert VT_{2}\right\rangle
\left\vert VT_{2}\right\rangle -\cos 2\theta _{2}\left\vert
HT_{2}\right\rangle \left\vert HT_{2}\right\rangle )\text{,}
\end{eqnarray}%
and the four bases are $\{\left\vert HT_{1}\right\rangle $,
$\left\vert VT_{1}\right\rangle $, $\left\vert HT_{2}\right\rangle
$, $\left\vert VT_{2}\right\rangle \}$ (If $\theta _{1}$ or
$\theta _{2}=0$, it reduces to $ 2\times 2$ dimensional entangled
state). To enhance the purity of two-photon states, we make
down-converted photons pass through the interference filter,
compensation crystal (CC) and enter into the single mode fiber.
The interference filter is centered at 780nm and its bandwidth is
3nm, which corresponds to $\tau =676$ fs for the correlation time
of down converted photons. In our experiment, the two-photon
coincidence window is $ \Delta t=$3ns. The visibility for
two-photon state and four-photon state are more than 96\% and 95\%
respectively\cite{Xu}, indicating the high purity of the photon
state.

\begin{figure}[tbh]
\begin{center}
\includegraphics[width= 4in]{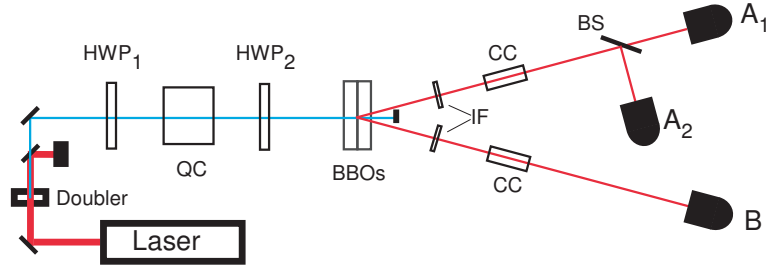}
\end{center}
\caption{(color online) Experimental settings. The pseudo-twofold
copy of the two-photon state is generated from the second order
PDC. } \label{fig2}
\end{figure}

It is worth mentioning the time scales in the experiment. $\Delta
T>\tau _{p}$, $\tau $ guarantees good time separation of the two
pulses so that the orthogonality condition $\left\langle
P_{i}T_{i}|P_{j}T_{j}\right\rangle =\delta _{P_{i}P_{j}}\delta
_{ij}(P_{i},P_{j}\in \{H,V\};i,j\in \{1,2\})$ holds. $\tau $,
$\Delta T\ll \Delta t$ makes the time separation undetectable
through the photon coincidence counts, so that quantum coherence
of $\left\vert \Psi _{c}\right\rangle $ can be observed.

We measure the single photon counting rate of $A_{1}$, $A_{2}$ and
the coincidence counting rate between $A_{1}$ and $A_{2}$ as
$N_{A_{1}}$, $N_{A_{2}}$ and $ N_{A_{1}A_{2}}$, respectively. Then
$P_{A_{1}(A_{2})}=N_{A_{1}(A_{2})}/f$ , $
P_{A_{1}A_{2}}=N_{A_{1}A_{2}}/f$ and
$K=\frac{fN_{A_{1}A_{2}}}{N_{A_{1}}N_{A_{2}}}-1$. 

\begin{figure}[tbh]
\begin{center}
\includegraphics[width= 4.5in]{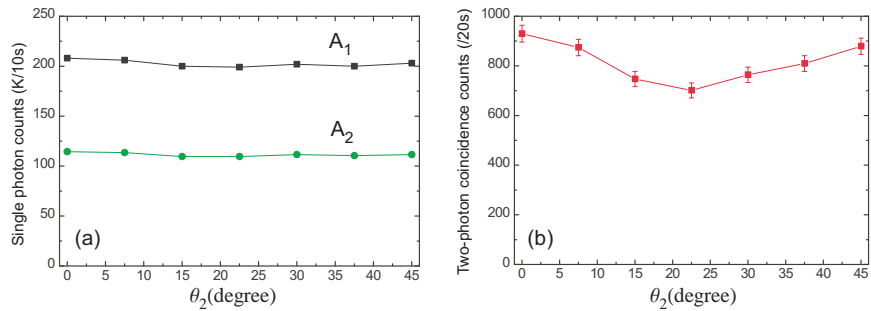}
\end{center}
\caption{(color online) Single photon Counts and two-photon coincidence
counts with $\protect\theta _{1}=$22.5$^{\circ }$.}
\label{fig3}
\end{figure}

\begin{figure}[tbh]
\begin{center}
\includegraphics[width= 4.5in]{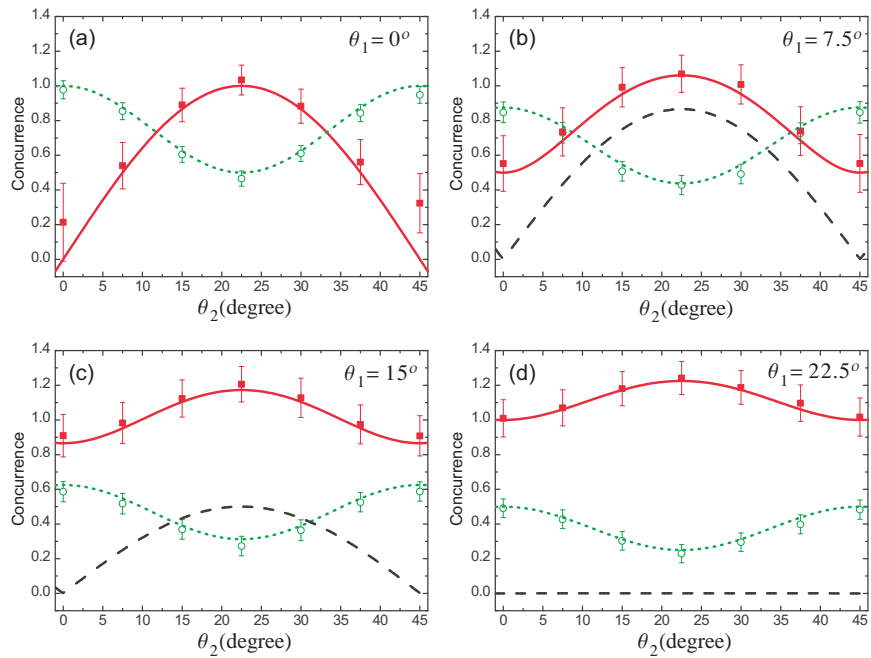}
\end{center}
\caption{(color online) Plot of entanglement measurement results
with different angles $\protect\theta _{1}$ and $\protect\theta
_{2}$. The green square points and dashed curves are the
experimental data and theoretic value of $K$. Red circle points
and curves are measured relative concurrences and their
theoretical values. The black dashed curves show the
sub-concurrence on polarization DOF after the time-energy DOF is
traced out.}
\end{figure}

Fig.3 is the experimental photon counts with $\theta
_{1}=$22.5$^{\circ }$. Single photon counts of A$_{1}$ and A$_{2}$
are indicated by black square and green circle points in Fig.3(a).
Fig.3(b) shows coincidence counts between A$_{1}$ and A$_{2}$. The
entanglement measurement result with different angles of the two
HWP $\theta _{1}$ and $\theta _{2}$ are shown in Fig.4. The green
open circle points are the data of $K$ and the green dotted curves
show the theoretical values with function of $K(\theta _{1},\theta
_{2})=(\cos ^{4}2\theta _{1}+\sin ^{4}2\theta _{1})(\cos
^{4}2\theta _{2}+\sin ^{4}2\theta _{2})$. The solid red square
points refer to relative \emph{I-concurrence} and the theoretical
values of $C(\theta _{1},\theta _{2})=\sqrt{ 2-2K(\theta
_{1},\theta _{2})}$ are illustrated with red solid curves.
Fig.4(a) shows the experimental results for $\theta
_{1}=$0$^{\circ }$, which corresponds to the case of $2\times 2$
entangled state. When $\theta _{2}=$ 22.5$^{\circ }$, it becomes
the maximally entangled (Bell) state and the \emph{I-concurrence}
reaches 1.03$\pm $0.09. While $\theta _{2}=$0$^{\circ }$ or 45$
^{\circ }$, it becomes a product state with minimal entanglement.
Fig.4(b), (c), and (d) depict the $4\times 4$ dimensional
entangled state with $\theta _{1}=$7.5$^{\circ }$, 15$^{\circ }$,
and 22.5$^{\circ }$, respectively. The $ 4\times 4$ maximally
entangled state can be achieved when both angles of $ \theta _{1}$
and $\theta _{2}$ are set to 22.5$^{\circ }$. The measured
\emph{I-concurrence} for this state is 1.24$\pm $0.09, whereas the
theoretical value is $\sqrt{6}/2$. When $\theta _{2}=$0$^{\circ }$
or 45$^{\circ }$, the states are reduced to $2\times 2$ dimension
again. The experimental results agree with the theoretical values
well within the experimental errors from photon counts variance.
Moreover, the experimental data shows the \emph{I-concurrence} $C$
is always no less than the sub-concurrence $C_{12}=2|(\cos
^{2}2\theta _{1}-\sin ^{2}2\theta _{1})\cos 2\theta _{2}\sin
2\theta _{2})|$ , the polarization-dependent concurrence when
time-energy DOF is traced out. $C_{12}$ is plotted as the black
dashed curve in Figs.4(b), (c), and (d).

\section{Discussions and Conclusions}
In our experiment, $K$ is always a little less than the
theoretical value of $Tr\rho _{A}^{2}$. It is likely because there
are other DOFs besides the polarization and time-energy DOF
involved in our experiment. It may be the frequency DOF, despite
the narrow frequency filters used to improve the purity of the
two-photon state \cite{Ou2}. When these additional DOFs are
present, the photon state will be that of a higher dimensional
system. Generally, $K$ less than the theoretical value for the
maximally entangled states indicates there are other dimensions
not under consideration. Therefore, our scheme could act as an
effective method to detect additional DOF whether it is entangled
with the main DOF or not.

In conclusion, we experimentally determine the entanglement
measure of two-partite pure photon state with an equivalent
symmetric projection measurement for the two-fold copy of single
subsystem. We find the \emph{I-concurrence} of entangled states
produced in 1-order optical PDC can be obtained by measuring
entangled states produced in 2-order optical PDC. Our method, for
the first time, takes advantage of the indistinguishability and
the bunching effect of photons to measure the entanglement of
bi-photon pure states, which is suitable for application in
optical PDC process to determine the entanglement of
high-dimensional bipartite pure states composed of other DOF of
photons.

\section{Acknowledgments}
This work was funded by National Fundamental Research Program No.
2006CB921900, NCET-04-0587, the Innovation funds from Chinese
Academy of Sciences, and National Natural Science Foundation of
China (Grant Nos. 60621064, 10574126).


\begin{thebibliography}{99}
\bibitem{efficiency} Here, we may neglect the single photon counts contribution from
the four-photon state due to the fact $\left\vert \eta \right\vert
\ll 1$.
\end{thebibliography}
\end{document}